\documentclass[aps,pre,twocolumn,amssymb,reprint,superscriptaddress]{revtex4}
\usepackage{graphicx,amsmath}
\usepackage{color,pifont}

\definecolor{dgreen}{RGB}{0,160,0}

\begin{document}
\title{Brownian particles on rough substrates: Relation between the intermediate subdiffusion and the asymptotic long-time diffusion}
\author{Richard D.L. Hanes}
\affiliation{Condensed Matter Physics Laboratory, Heinrich Heine University, D-40225 D\"usseldorf, Germany}
\author{Michael Schmiedeberg}
\email{schmiedeberg@thphy.uni-duesseldorf.de}
\affiliation{Institute for Theoretical Physics 2: Soft Matter, Heinrich Heine University, D-40225 D\"usseldorf, Germany}
\affiliation{Department of Physics, University of Osnabr\"uck, D-49076 Osnabr\"uck, Germany}
\author{Stefan U. Egelhaaf}
\email{stefan.egelhaaf@uni-duesseldorf.de}
\affiliation{Condensed Matter Physics Laboratory, Heinrich Heine University, D-40225 D\"usseldorf, Germany}

\begin{abstract}
Brownian particles in random potentials show an extended regime of subdiffusive dynamics at intermediate times. The asymptotic diffusive behavior is often established at very long times and thus cannot be accessed in experiments or simulations. For the case of one-dimensional random potentials with Gaussian distributed energies, we present a detailed analysis of experimental and simulation data. It is shown that the asymptotic long-time diffusion coefficient can be related to the behavior at intermediate times, namely the minimum of the exponent that characterizes subdiffusion and hence corresponds to the maximum degree of subdiffusion. As a consequence, investigating only the dynamics at intermediate times is sufficient to predict the order of magnitude of the long-time diffusion coefficient and the timescale at which the crossover from subdiffusion to diffusion occurs, i.e.~when the long-time diffusive regime and hence thermal equilibrium is established. 

\end{abstract}

%82.70.Dd:colloids 
%05.40.Jc Brownian motion
%05.60.-k Transport processes 

\maketitle

\section{Introduction} 
\label{sec:intro}

The motion of Brownian particles on a rough surface or in a random external potential exhibits different dynamical regimes. If the external potential lacks a lower limit and the mean or second moment of the potential energy do not exist, the motion remains subdiffusive even in the asymptotic long-time limit \cite{honkonen89,romero98,sancho04,lacasta04}. However, if the first and second moments of the potential energy are well defined and finite, the motion becomes diffusive in the long-time limit, i.e., the mean square displacement $\left\langle r^2(t) \right\rangle$ is proportional to the time $t$ for $t\rightarrow \infty$. Nevertheless, at intermediate times an extended subdiffusive regime usually exists where $\left\langle r^2(t) \right\rangle\propto t^{\nu(t)}$ with the exponent $\nu(t)<1$ (for reviews see, e.g., \cite{haus87,bouchaud90,bouchaud90b}). Colloidal model systems can be used to systematically study the intermediate and long-time dynamics experimentally \cite{tierno10,tierno12,hanes12, evers13, evers13review, ma13,volpe13} or with simulations \cite{romero98,sancho04,lacasta04,schmiedeberg07,emary12,hanes12b}. The crossover from the intermediate subdiffusion to the long-time diffusion occurs at progressively longer times as the roughness of the surface or the barriers of the potential are increased. As a consequence, the asymptotic long-time diffusive regime is often inaccessible in experiments or simulations.

The properties of the asymptotic long-time dynamics, such as the asymptotic diffusion coefficient $D_\infty$, can be theoretically derived within various models, e.g., diffusion models with rough potentials \cite{zwanzig88}, hopping, transition rate or random trap models \cite{haus82,derrida83,schmiedeberg07}, random barrier methods \cite{bernasconi79,jack09,novikov11}, or continuous-time random walks \cite{scher73,metzler00}. However, a comprehensive theoretical description of the intermediate subdiffusive regime is still lacking.

This is despite the existence of intermediate subdiffusive regimes in the dynamics of many systems. In addition to the already mentioned Brownian motion in random and also regular potentials \cite{volpe13,tierno10, tierno12,loewen08, emary12, zwanzig88,reimann02, euan12, schmiedeberg07,jenkins08b,dalle11,hanes12,hanes12b,evers13,evers13review,ma13} or on rough surfaces \cite{naumovets05, barth00, sengupta05}, subdiffusion is also observed when particles move in confinement \cite{wei00}, in inhomogeneous media (e.g.~with fixed obstacles as in a Lorentz gas \cite{hoefling06}, in porous gels \cite{dickson96} or cells \cite{tolic04, weiss04, hoefling13}), in materials with defects (e.g.~zeolites \cite{chen00} or charge carriers in a conductor with impurities \cite{bystroem50, heuer05}), or between magnetic domains \cite{tierno10, tierno12}. Intermediate subdiffusive regimes also occur in dense suspensions close to freezing \cite{indrani94} and glasses \cite{goetze99,debenedetti01,angell95}, where subdiffusion is due to particles being trapped in the cage of neighbors and has been described by potential energy landscape models \cite{debenedetti01,angell95,heuer08, sciortino05} that are similar to random trap models. Furthermore, in biological systems a similar phenomenon, termed crowding, can occur at large densities \cite{weiss04, hoefling13}. Energy landscapes have also been applied in the context of protein folding \cite{durbin96, dill97} and the behavior of RNA, proteins and transmembrane helices \cite{hyeon03, janovjak07}, where random energy landscapes with a Gaussian distribution of energy levels of width ${\cal{O}}(k_{\mathrm{B}}T)$, where $k_{\mathrm{B}}T$ is the thermal energy, seem to be relevant. In these examples, diffusion in a random potential energy landscape might represent a crude approximation only, but nevertheless often provides a useful initial description of the effect of disorder on the dynamics \cite{bouchaud90b,wolynes92}.

Here, experiments and simulations are performed to investigate the dynamics of a colloidal particle in an external potential, namely a one-dimensional random potential whose potential values are distributed according to a Gaussian of width $\epsilon$. We determine the asymptotic long-time diffusion coefficient $D_{\infty}$ and the time scale $\tau_\infty$ associated with the crossover from subdiffusion to asymptotic long-time diffusion, as well as the exponent $\nu(t)$ that characterizes the intermediate sub\-diffusive regime. We find that $D_{\infty}\propto \tau_\infty^{-1}\propto\exp\left[-(\epsilon/(k_{\mathrm{B}}T))^2\right]$ in agreement with theoretical predictions \cite{zwanzig88,schmiedeberg07} and that the minimum of $\nu(t)$ approximately follows $\nu_{\textnormal{min}}=\exp\left[-c\,\epsilon/(k_{\mathrm{B}}T)\right]$ with a constant $c$.  Using these relations, we demonstrate that, if one obtains $\nu_{\textnormal{min}}$ in the intermediate regime and $D_{\infty}$ for a few (possibly small) $\epsilon$, the order of magnitude of $D_{\infty}$ and $\tau_\infty$ can be estimated even for large $\epsilon$, i.e.~for conditions where it is difficult or even impossible to reach the asymptotic regime.

The article is structured as follows: In Sec.~\ref{sec:model} we introduce the model system and describe details of the experiments and simulations. The results are presented in Sec.~\ref{sec:results}. In Sec.~\ref{sec:prediction} we demonstrate how the long-time diffusion coefficient $D_\infty$ and the crossover time $\tau_\infty$ can be predicted even for large $\epsilon$. Finally, we conclude in Sec.~\ref{sec:conclusions}.

\section{System} 
\label{sec:model}

\subsection{Experiment}

The sample consisted of a suspensions of colloidal spheres made from polystyrene with sulfonated chain ends (Interfacial Dynamics Corporation) with radius $R = 1.4\,\mu$m in heavy water. The suspension was dilute with an area fraction of the quasi two-dimensional (creamed) particle layer of less than 0.05 to minimise particle--particle interactions. The sample was kept in a cell constructed from microscope slides and cover slips which were thoroughly cleaned to reduce sticking of particles to the glass surfaces; two cover slips were used as spacers with a horizontal gap between them and a third cover slip on top resulting in a narrow capillary \cite{jenkins08}.

An external potential was imposed on the polarizable particles by exposing them to a light field \cite{ashkin97,molloy02}. The light field was created using a laser with a wavelength of 532 nm (Ventus 532-1500, Laser Quantum) and a spatial light modulator (Holoeye 2500-LCR) \cite{hanes09, hanes12}. The light fields consisted of rings of high average intensity with random intensity fluctuations. Different realizations of the fluctuations were created, all of them leading to a random potential exerted on the particles with the distribution of energy levels following a Gaussian distribution with standard deviation, or degree of roughness, $\epsilon$. The roughness of the potential, $\epsilon$, is controlled via the laser intensity.

The sample was imaged using a Nikon Eclipse 2000-U inverted microscope with a Plan APO VC Oil $60\times$ objective. Micrographs were recorded with a CMOS camera (PL-B742F, Pixelink). From the time series of micrographs, particle coordinates were extracted \cite{crocker96} and subsequently trajectories $x_i(t)$ of the individual particles $i$ determined. Details of the experiments and data analysis are given in \cite{hanes12}.

\subsection{Simulations} 

In the simulations, first random potential values $\overline{U}(x)$ are drawn from a Gaussian distribution with standard deviation $\overline{\epsilon}$. The resulting $\overline{U}(x)$ correspond to the spatially varying laser intensity. A convolution of $\overline{U}(x)$ with the volume of the spherical particle results in the effective potential $U(x)$ as felt by a point-like particle at position $x$ \cite{hanes12b}. The effective potential $U(x)$ has Gaussian-distributed potential values with standard deviation $\epsilon$. Initially, the particle was randomly positioned in the potential, corresponding to an instantaneous quench of the system.
At each time step, the particle attempts to move a distance $x_\mathrm{s}=R/32$ with the direction chosen randomly. The move is executed if the potential energy of the new position is smaller than the current potential energy. Otherwise, the move is accepted with a probability $\exp\left[-\Delta U/(k_{\mathrm{B}}T)\right]$, where $\Delta U$ is the difference between the potential values at the new and current positions. For the determination of the different parameters characterizing the particle dynamics (Sec.~\ref{sec:msd}), $5000$ individual runs were averaged. Times are normalized by the Brownian time $\tau_{\mathrm{B}}=R^2/(2D_0)$, which is the time that a particle requires to diffuse its own radius $R$ in free diffusion. The free diffusion coefficient $D_0$ is obtained for $\epsilon\rightarrow 0$. Details of the simulations are described in \cite{hanes12b}.

\section{Results} 
\label{sec:results}

\subsection{Mean square displacement, diffusion coefficient, and degree of subdiffusion} 
\label{sec:msd}

\begin{figure}
\includegraphics[width=0.98\columnwidth]{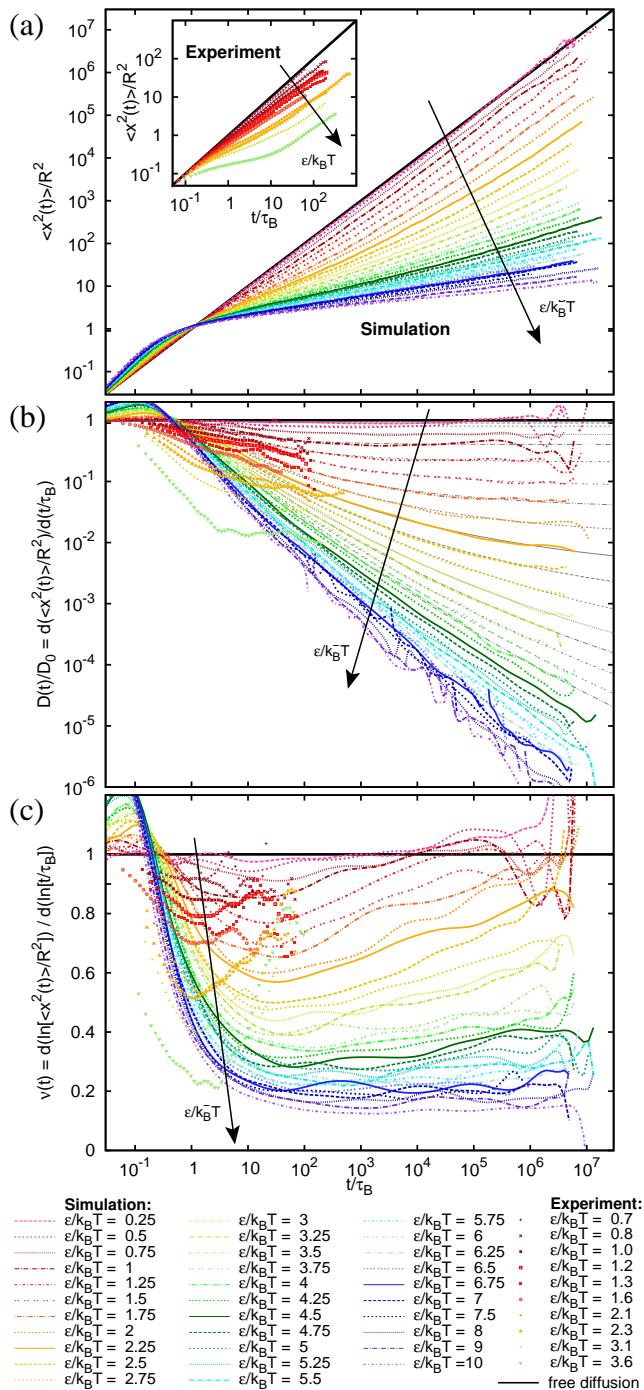}
\caption{(Color online) Time dependence of the (a) mean square displacement $\left\langle x^2(t)\right\rangle$ in units of the radius $R$, (b) diffusion coefficient $D$ in units of the free diffusion coefficient $D_0=R^2/(2\tau_{\mathrm{B}})$, and (c) exponent $\nu(t)$ of the mean square displacement as in $\left\langle x^2(t)\right\rangle\propto t^{\nu(t)}$ for different degrees of roughness of the potential, $\epsilon/k_{\mathrm{B}}T$ (as indicated). Time $t$ in units of the Brownian time $\tau_{\mathrm{B}}$. Simulation results are represented by lines in the main figure, experiments by symbols (in (a) in the inset). Note that the simulation and experimental results are averaged differently (see text). Fits to $D(t)/D_0$ from simulations are represented by thin black lines. For comparison, free diffusion is indicated by solid black lines with (a) $\left\langle x^2(t)\right\rangle=2D_0 t$, (b) $D/D_0=1$, and (c) $\nu=1$.
\label{fig:msd}}
\end{figure}

Based on the particle trajectories $x_i(t)$, the mean square displacement is calculated according to
\begin{equation}
\left\langle x^2(t)\right\rangle = \left\langle \left[x_i(t_0+t)-x_i(t_0)\right]^2 \right\rangle
\end{equation}
where the average is taken over different particles $i$ and, to improve the statistics of the experimental results, waiting times $t_0$. In contrast, in the simulations the average is only taken over different particles $i$, but not $t_0$, which is set to $t_0=0$ corresponding to the time when the system is quenched. The mean square displacement $\left\langle x^2(t)\right\rangle$ as a function of delay time $t$ shows a strong dependence on the standard variation $\epsilon$ of the distribution of potential values $U(x)$, which is a measure for the roughness of the potential (Fig.~\ref{fig:msd}(a)). For vanishing $\epsilon=0$ (black solid line), free diffusion is observed. For $\epsilon>0$, subdiffusive dynamics occurs at intermediate times. It becomes more pronounced and extends to longer times as $\epsilon$ increases. For long times, the dynamics becomes diffusive again, although with a reduced diffusion coefficient $D_\infty$. The crossover from intermediate subdiffusion to the asymptotic diffusive regime occurs at increasingly longer times $\tau_\infty$ as $\epsilon$ increases. For very large $\epsilon$, the asymptotic regime is not reached within the observation time.

At very short times, superdiffusion is observed in the simulations. This is due to the particle being driven from its initial quenched position to the closest (most likely local) minimum. As $\epsilon$ increases, the slopes become steeper and hence the particle is more strongly driven, reflected in a more pronounced superdiffusion. In the experimental $\left \langle x^2(t) \right \rangle$ (Fig.~\ref{fig:msd}(a), inset), the initial superdiffusion is masked due to the average over waiting times $t_0$. The average allows the behavior at later times to contribute to $\left \langle x^2(t) \right \rangle$ and hence results in only a small weight of the initial superdiffusive regime. (Note that for the simulations $t_0=0$.) The averaging over $t_0$ has a further consequence: The system is initially quenched and evolves toward an occupation of the energy values following a Boltzmann distribution. This implies an increasing occupation of deep minima. The escape from deeper minima takes longer and hence results in slower dynamics. The averaging, via the inclusion of later times with their slower dynamics, thus results in apparently enhanced subdiffusion. This is indeed observed in the experimental $t_0$-averaged $\left \langle x^2(t) \right \rangle$ (see also Figs.~11, 12 in \cite{hanes12b}). Nevertheless, since the asymptotic long-time regime is only reached after the system equilibrated, the long-time limit is not affected by the averaging over $t_0$.

The time-dependent diffusion coefficient $D(t)$ can be defined as the derivative of $\left \langle x^2(t) \right \rangle$:
\begin{equation}
D(t) = \frac{{\mathrm{d}}\left \langle x^2(t) \right \rangle} {2 \; {\mathrm{d}}t} \;\; .
\end{equation}
Fig.~\ref{fig:msd}(b) shows $D(t)$ in units of the free diffusion coefficient $D_0$ as a function of the delay time $t$ for different degrees of roughness of the potential, $\epsilon$, as obtained from simulations (lines) and experiments (symbols). In case of free diffusion, that is $\epsilon=0$, $D(t)/D_0=1$. In the presence of a random potential, $D(t)$ monotonically decreases at intermediate times until, in the asymptotic regime, diffusive behavior is recovered with a constant asymptotic long-time diffusion coefficient $D_{\infty}$. With increasing $\epsilon$, the decrease of $D(t)$ becomes more pronounced and the asymptotic regime is reached at increasingly longer times. In the log-log-plot, the approach of $\log{(D(t)/D_0)}$ towards the asymptotic value $\log{(D_\infty/D_0)}$ can be described by an exponential function:
\begin{equation}
\log_{10}\left(\frac{D(t)}{D_0}\right)=\log_{10}\left(\frac{D_{\infty}}{D_0}\right)\left[1+a \exp\left(-t/\tau_{\infty}\right)\right],
\label{eq:fitDt}
\end{equation}
with a fit constant $a$.
In Fig.~\ref{fig:msd}(b) thin black lines indicate fits to the simulation data. The fits are used to determine $D_{\infty}$ and $\tau_{\infty}$ even if the long-time limit is not reached within the simulation time. Since the experimental data are averaged over $t_0$, they contain a significant contribution from the dynamics at late times and hence of the system closer to equilibrium where the particle tends to occupy lower energy values. This leads to a sharper decrease of $D(t)/D_0$ at short and intermediate times, but, in the asymptotic long-time limit, to the same $D_\infty/D_0$ \cite{hanes12b}.

The mean square displacement $\left\langle x^2(t)\right\rangle$ at delay time $t$ can be expressed as a power law $\left\langle x^2(t)\right\rangle\propto t^{\nu(t)}$. The time-dependent exponent $\nu(t)$ can be calculated using
\begin{equation}
\nu(t)=\frac{{\mathrm{d}} \log_{10} \left(\left\langle x^2(t)\right\rangle/R^2\right)}{{\mathrm{d}}\log_{10} \left(t/\tau_{\mathrm{B}}\right)}.
\end{equation}
In Fig.\ \ref{fig:msd}(c) the exponent $\nu(t)$ is shown as a function of the delay time $t$. In the absence of an external potential, i.e.~$\epsilon=0$, free diffusion with $\nu(t)=1$ is observed. For $\epsilon > 0$, $\nu(t)<1$ and thus subdiffusion occurs. The sharp decrease of $\nu(t)$ is due to the particle being trapped in a local minimum with the trapping becoming more efficient with increasing $\epsilon$. In contrast, the crossover from subdiffusion to asymptotic diffusion, indicated by $\nu(t)$ approaching $1$, is very slow and occurs at increasingly longer times as $\epsilon$ increases. For the largest $\epsilon$, it cannot be determined within our observation window. For diffusion to be re-established, the particle has to escape also deep minima and hence cross large barriers. This requires a very long time which depends on the range of barrier heights, i.e.~$\epsilon$. Furthermore, as $\epsilon$ increases, the minimum in $\nu(t)$, $\nu_{\mathrm{min}}$, decreases, which will be analyzed in Sec.~\ref{sec:il}. Due to the $t_0$ averaging, the experimental $\nu(t)$ also contain a contribution from the dynamics at later times, when the particle already escaped the minima, and thus increase toward 1 earlier. The value of $\nu_{\mathrm{min}}$ is, however, hardly affected by the $t_0$ averaging as shown below.

\subsection{Dynamics at intermediate times and in the asymptotic long-time limit}
\label{sec:il}

\begin{figure}
\includegraphics[width=\columnwidth]{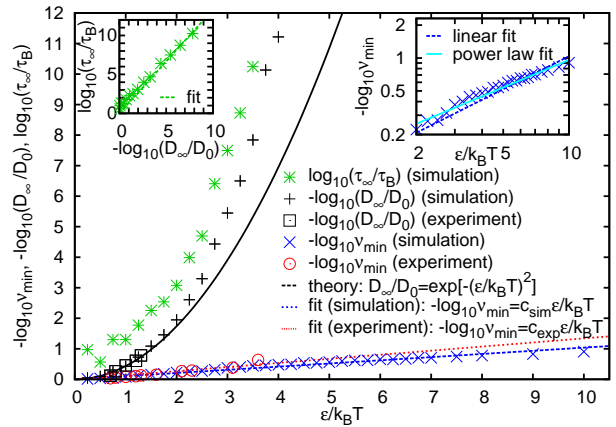}
\caption{(Color online) Negative logarithm of the normalized asymptotic diffusion coefficient $D_{\infty}/D_0$ as a function of the degree of roughness of the potential, $\epsilon/k_{\mathrm{B}}T$. $D_{\infty}/D_0$ is obtained by theoretical predictions (Eq.~\ref{eq:dinf}, black solid line) and fits to $D(t)/D_0$ (Fig.~\ref{fig:msd}(b)) which was determined in simulations (+) and experiments ($\boxdot$). The logarithm of the normalized timescale $\tau_{\infty}/\tau_{\mathrm{B}}$, assosiated with the crossover from subdiffusion to asymptotic diffusion, is shown for the simulation results (\textcolor{dgreen}{\ding{83}}). The negative logarithm of the minimal value of the exponent, $\nu_{\textnormal{min}}$, is obtained from $\nu(t)$ (Fig.~\ref{fig:msd}(c)) determined in simulations (\textcolor{blue}{$\times$}) and experiments (\textcolor{red}{$\odot$}). Straight lines are fitted to $\log{(\nu_{\textnormal{min}}(\epsilon))}$ (simulations: dashed blue line, experiments: dotted red line). The inset on the right shows $-\log{(\nu_{\textnormal{min}})}$ from simulations together with a linear fit as in the main figure (dashed blue line) and a power law fit with exponent $0.83$ (solid cyan line). The inset on the left shows $\log\left(\tau_{\infty}/\tau_{\mathrm{B}}\right)$ as a function of $-\log\left(D_{\infty}/D_0\right)$. The broken green line indicates a linear dependence $\tau_{\infty}/\tau_{\mathrm{B}}=fD_{\infty}/D_0$ (Eq. (\ref{eq:tau})) where $f\approx 1.33$ is obtained by fitting.}
\label{fig:vergl}
\end{figure}

The asymptotic long-time diffusion coefficient $D_\infty$ of diffusion in a one-dimensional random potential was calculated to be \cite{zwanzig88}
\begin{equation}
\label{eq:dinf}
\frac{D_{\infty}}{D_0}=\exp\left[-\left\{\frac{\epsilon}{k_BT}\right\}^2\right].
\end{equation}
The same relation was also derived from transition rate models (see, e.g., \cite{haus87}) and continuous-time random walks with transition rates calculated according to Kramers' formula \cite{haenggi90}. For small $\epsilon$, the agreement between the theory and our experiments and simulations is very good, while for large $\epsilon$ the simulations lead to smaller values of $D_{\infty}$, i.e.~larger values of $-\log_{10}(D_\infty/D_0)$, than expected from theory (Fig.~\ref{fig:vergl}). This is due to the fact that, for large $\epsilon$, the asymptotic diffusive regime was not reached within the simulation time and hence $D_{\infty}$ was determined by extrapolating the time dependent diffusion coefficient $D(t)$ (Fig.~\ref{fig:msd}(b), thin black lines), which seems to systematically underestimate $D_{\infty}$.

The timescale $\tau_\infty$ quantifies when the crossover from subdiffusion to asymptotic diffusion occurs, i.e.~the asymptotic long-time regime is established. From the fits to the simulation data (Fig.~\ref{fig:msd}), the $\epsilon$ dependence of $\tau_\infty$ is extracted (Fig.~\ref{fig:vergl}). Note that the value of $\tau_\infty$ might depend on the (heuristic) fit equation used (Eq.~\ref{eq:fitDt}), for both, small $\epsilon$, where $D(t)/D_0$ shows only a weak time dependence, as well as large $\epsilon$, where a significant extrapolation is required. The timescale $\tau_\infty$ is predicted to follow \cite{schmiedeberg07}
\begin{equation}
\label{eq:tau}
\tau_\infty \approx f \tau_{\mathrm{B}} \frac{D_0}{D_{\infty}} \;\; ,
\end{equation}
where $f$ is a prefactor of order 1 that depends on the details of the potential. From a fit to our simulation results (Fig. \ref{fig:vergl}, left inset) we find $f\approx 1.33$. Based on $D_{\infty}$, thus $\tau_\infty$ can be estimated even if it cannot be extracted directly from the data. The crossover time $\tau_{\infty}$ is of special interest for many simulations and experiments since it characterizes the relaxation time required to reach thermal equilibrium.

The intermediate subdiffusive regime is characterized by a minimum of $\nu(t)$ (Fig.~\ref{fig:msd}(c)). The minimum $\nu_{\textnormal{min}}$ was determined as a function of $\epsilon$ (Fig.~\ref{fig:vergl}). With increasing $\epsilon$,  $\nu_{\textnormal{min}}$ decreases indicating the increasingly pronounced subdiffusion. The logarithm of $\nu_{\textnormal{min}}$ can be fitted by a linear function, namely
\begin{equation}
\nu_{\textnormal{min}} = \exp\left[-c\frac{\epsilon}{k_BT}\right]
\label{eq:nu}
\end{equation}
with a constant $c$. We find $c=c_{\textnormal{sim}}\approx 0.104$ for the simulation data and $c=c_{\textnormal{exp}}\approx 0.134$ in case of the experiments. A power law fit with exponent $0.83$ seems to be slightly more suitable than the linear fit (Fig.~\ref{fig:vergl}, right inset). However, for simplicity and because the difference is very small, the linear fit (Eq.~\ref{eq:nu}) will be used in the next section.

The time $\tau_{\textnormal{min}}$ at which the minimum occurs is very difficult to determine unambiguously due to the shallow minimum, especially for large $\epsilon$. We thus refrain from extracting $\tau_{\textnormal{min}}$.

\subsection{Predicting the asymptotic long-time dynamics based on the intermediate subdiffusion} 
 \label{sec:prediction}

\begin{figure}
\includegraphics[width=\columnwidth]{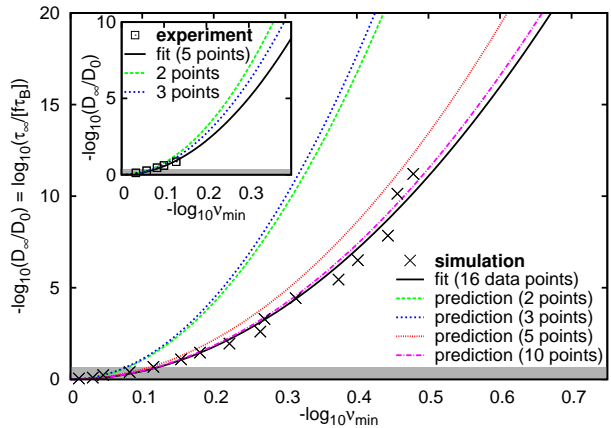}
\caption{(Color online) Negative logarithm of the normalized asymptotic diffusion coefficient $D_{\infty}/D_0$, which is related to the logarithm of the time characterizing the crossover to the asymptotic regime $\tau_\infty$, as a function of the negative logarithm of the minimum exponent $\nu_{\mathrm{min}}$ as obtained by simulations (main figure, $\times$) and experiments (inset, $\boxdot$). Lines are quadratic fits to all data (black solid line) or only the first few data points (broken lines). The grey area indicates the timescale which needs to be explored in experiments or simulations in order to predict the behavior shown in the entire figure.}
\label{fig:pred}
\end{figure}

In the previous section we have determined characteristic features of the intermediate subdiffusive regime, namely the minimum of the exponent $\nu_{\mathrm{min}}$, as well as of the asymptotic long-time regime, that is the asymptotic long-time diffusion coefficient $D_\infty$ and crossover time $\tau_\infty$. These parameters only depend on the degree of the roughness of the potential, $\epsilon$ (Eqs.~\ref{eq:dinf}, \ref{eq:tau}, \ref{eq:nu}). They can thus be related to each other; $-\log_{10} D_{\infty}$ shows a quadratic dependence on $-\log_{10} \nu_{\textnormal{min}}$ (Fig.~\ref{fig:pred}). Interestingly, the first few data points obtained for small $\epsilon$ are sufficient to reliably determine the only fit parameter, $c$ (Fig.~\ref{fig:pred}, broken lines). While, in the case of simulations, fits to the first two or three points lead to an underestimate of $D_\infty$ (i.e.~overestimate of $-\log_{10} D_{\infty}$), fits to the first five points from the simulations and only two points from the experiments, respectively, result in good estimates for all data points including those at the highest $-\log_{10} \nu_{\textnormal{min}}$ and thus large $\epsilon$. (Note that the required number of points depends on their $\nu_{\mathrm{min}}$ values not their determination by simulations or experiments.) The limited range of data, and thus timescales, used to reliably determine $c$ is highlighted by the grey area in Fig.~\ref{fig:pred}.

As a consequence, if in an experiment or simulation $\nu_{\textnormal{min}}$ and $D_{\infty}$ can be measured for a few, possibly small, $\epsilon$, i.e., on timescales that are indicated by the grey area in Fig.~\ref{fig:pred}, a quadratic fit to the logarithms of these data can provide the fit parameter $c$ and hence a relation between the parameters describing the asymptotic long-time behaviour, $D_{\infty}$ and $\tau_\infty$, and the one characterizing the intermediate regime, $\nu_{\textnormal{min}}$. Then a determination of  $\nu_{\textnormal{min}}$, which can be performed at intermediate times, will provide an estimate of the asymptotic long-time behavior, namely $D_\infty$ and $\tau_\infty$. 
Importantly, the duration of simulations and experiments required to obtain $\nu_{\textnormal{min}}$ is much shorter, often by many orders of magnitude, than required to determine the long-time dynamics (Fig.~\ref{fig:msd}), which is given by the crossover time to the asymptotic regime, $\tau_\infty$ (Fig.~\ref{fig:vergl}). Moreover, in $t_0$ averaged data, the minimum in $\nu(t)$ occurs earlier.  Therefore, even if thermal equilibrium is not reached within the simulations or experiments, the timescale on which the relaxation will take place can be estimated. Hence $D_\infty$ and $\tau_\infty$ can be estimated even for very rough substrates or potentials without having to perform long simulations or experiments. Furthermore, the roughness $\epsilon$ of the surface or potential does not need to be known to obtain an estimate of $D_{\infty}$ and $\tau_\infty$. 

Moreover, in experiments or simulations with particles on rough surfaces or in random potentials, the roughness $\epsilon$ can often be varied but not quantified. If $\epsilon$ cannot be determined, the relationship between $\nu_{\textnormal{min}}$ and $D_{\infty}$ can be exploited to obtain $\epsilon$. Determining a few sets of $\nu_{\textnormal{min}}$ and $D_{\infty}$, possibly on short timescales, i.e.~for small $\epsilon$, allows for the determination of $c$. Subsequently, $D_{\infty}$ and $\epsilon$ can be predicted as a function of $\nu_{\textnormal{min}}$.

\section{Conclusions} 
\label{sec:conclusions}

The motion of individual colloidal particles was studied in random potentials using simulations and experiments. We in particular investigated the dynamics in the intermediate subdiffusive regime and in the asymptotic long-time regime, where the motion again is diffusive. The behavior at very long times, namely the asymptotic long-time diffusion coefficient $D_\infty$ and the crossover time from subdiffusion to diffusion $\tau_\infty$, was related to the characteristic feature at intermediate times, that is the minimum in the exponent $\nu_{\mathrm{min}}$, which quantifies the degree of subdiffusion. As predicted by theory \cite{zwanzig88,schmiedeberg07}, the logarithms of $D_\infty/D_0$ and $\tau_\infty/\tau_{\mathrm{B}}$ are quadratic functions of $\epsilon$, while the logarithm of $\nu_{\textnormal{min}}$ was found to be approximately a linear function of $\epsilon$. This allowed us to relate $D_\infty$ and $\tau_\infty$ to $\nu_{\textnormal{min}}$ (Fig.~\ref{fig:pred}) and thus the properties of the asymptotic long-time dynamics to the intermediate dynamics.

In the case of very rough surfaces or potentials, the asymptotic diffusive regime occurs at very long times. It thus often is not accessible in experiments and simulations and hence $D_\infty$ and $\tau_\infty$ cannot be measured. However, we have demonstrated that if one determines $\nu_{\textnormal{min}}$, which requires only an investigation at intermediate times, and a few values of $D_{\infty}$, possibly at a small degree of roughness $\epsilon$, then $D_{\infty}$ and $\tau_\infty$ can be predicted even for rough substrates and potentials, i.e.~large $\epsilon$. Our method can therefore be used to estimate, based on relatively short measurements, the asymptotic long-time diffusion coefficient $D_\infty$ and the crossover time $\tau_\infty$, and hence the time required to relax and reach thermal equilibrium without knowledge of $\epsilon$. Thus, the characteristic features of the asymptotic long-time dynamics can be determined based on measurements in the intermediate regime, i.e.~even if thermal equilibrium is not reached within the time of the experiment or simulation. 

We thank A. Heuer (M\"unster) as well as J. Bewerunge, F. Evers, and C. Zunke (D\"usseldorf) for very helpful discussions. We gratefully acknowledge support by the Deutsche Forschungsgemeinschaft (DFG) through the SFB-TR6 (project C7) and the International Helmholtz Research School `BioSoft'. M.S. also acknowledges support by the DFG within the Emmy Noether program (Schm 2657/2).


\begin{thebibliography}{99}

\bibitem{honkonen89}
J. Honkonen, Y.M. Pis'mak, J. Phys. A {\bf{22}}, L899 (1989).

\bibitem{romero98}
A.H. Romero, J.M. Sancho, Phys. Rev. E {\bf{58}}, 2833 (1998).

\bibitem{sancho04}
J.M. Sancho, A.M. Lacasta, K. Lindenberg, I.M. Sokolov, A.H. Romero, Phys. Rev. Lett. {\bf{92}}, 250601 (2004).

\bibitem{lacasta04}
A.M. Lacasta, J.M. Sancho, A.H. Romero, I.M. Sokolov, K. Lindenberg, Phys. Rev. E {\bf{70}}, 051104 (2004).

\bibitem{haus87}
J.W. Haus, K.W. Kehr, Phys. Rep. {\bf{150}}, 263 (1987).

\bibitem{bouchaud90}
J.-P. Bouchaud and A. Georges, Phys. Rep. {\bf{195}}, 127 (1990).

\bibitem{bouchaud90b}
J.-P. Bouchaud, A. Comtet, A. Georges, and P.L. Doussal, Ann. Phys. {\bf{201}}, 285 (1990).

\bibitem{tierno10}
P. Tierno, P. Reimann, T.H. Johansen, and F. Sagu{\'e}s, Phys. Rev. Lett. {\bf{105}}, 230602 (2010).

\bibitem{tierno12}
P. Tierno, F. Sagu{\'e}s, T.H. Johansen, and I.M. Sokolov, Phys. Rev. Lett. {\bf{109}}, 070601 (2012).

\bibitem{hanes12}
R.D.L. Hanes, C. Dalle-Ferrier, M. Schmiedeberg, M.C. Jenkins, and S.U. Egelhaaf, Soft Matter {\bf{8}}, 2714 (2012).

\bibitem{evers13}
F. Evers, C. Zunke, R.D.L. Hanes, J. Bewerunge, I. Ladadwa, A. Heuer, and S.U. Egelhaaf, Phys. Rev. E {\bf{88}}, 022125 (2013).

\bibitem{evers13review}
F. Evers, R.D.L. Hanes, C. Zunke, R.F. Capellmann, J. Bewerunge, C. Dalle-Ferrier, M. C. Jenkins, I. Ladadwa, A. Heuer, R. Casta{\~n}eda-Priego, and S.U. Egelhaaf, Eur. Phys. J. ST, accepted (2013).

\bibitem{ma13}
X. Ma, P. Lai, and P. Tong, Soft Matter, {\bf{9}} 8826 (2013).

\bibitem{volpe13}
G. Volpe, G. Volpe, and S. Gigan, {\it{Brownian Motion in a Speckle Light Field: Tunable Anomalous Diffusion and Deterministic Optical Manipulation}}, arXiv:1304.1433 (2013).

\bibitem{schmiedeberg07}
M. Schmiedeberg, J. Roth, and H. Stark, Eur. Phys. J. E {\bf{24}}, 367 (2007).

\bibitem{emary12}
C. Emary, R. Gernert, and S.H.L. Klapp, Phys. Rev. E {\bf{86}}, 061135 (2012).

\bibitem{hanes12b}
R.D.L. Hanes and S.U. Egelhaaf, J. Phys.: Condens. Matter {\bf{24}}, 464116 (2012).

\bibitem{zwanzig88}
R. Zwanzig, Proc. Natl Acad. Sci. {\bf{85}}, 2029 (1988).

\bibitem{haus82}
J.W. Haus, K.W. Kehr, and J.W. Lyklema, Phys. Rev. B {\bf{25}}, 2905 (1982).

\bibitem{derrida83}
B. Derrida, J. Stat. Phys. {\bf{31}}, 433 (1983).

\bibitem{bernasconi79}
J. Bernasconi, H.U. Beyeler, S. Str\"assler, and S. Alexander, Phys. Rev. Lett. {\bf{42}}, 819 (1979).

\bibitem{jack09}
R.L. Jack and P. Sollich, J. Stat. Mech. P11011 (2009).

\bibitem{novikov11}
D.S. Novikov, E. Fieremans, J.H. Jensen, and J.A. Helpern, Nature Phys. {\bf{7}}, 508 (2011).

\bibitem{scher73}
H. Scher and M. Lax, Phys. Rev. B {\bf{7}}, 4491 (1973).

\bibitem{metzler00}
R. Metzler and J. Klafter, Phys. Rep. {\bf{339}}, 1 (2000).

\bibitem{loewen08}
H. L\"owen, J. Phys.: Condens. Matter {\bf{20}}, 404201 (2008).

\bibitem{euan12}
E.C. Euan-Diaz, V.R. Misko, F.M. Peeters, S. Herrera-Velarde, and R. Casta{\~n}eda-Priego, Phys. Rev. E {\bf{86}}, 031123 (2012).
\bibitem{reimann02}
P. Reimann, C. Van den Broeck, H. Linke, P. H{\"a}nggi, J. M. Rubi, and A. P{\'e}rez-Madrid, Phys. Rev. E {\bf{65}}, 031104 (2002).

\bibitem{dalle11}
C. Dalle-Ferrier, M. Kr\"uger, R.D.L. Hanes, S. Walta, M.C. Jenkins, and S.U. Egelhaaf, Soft Matter {\bf{7}}, 2064 (2011).

\bibitem{jenkins08b}
M.C. Jenkins and S.U. Egelhaaf, J. Phys.: Condens. Matter {\bf{20}}, 404220 (2008).

\bibitem{naumovets05}
A. Naumovets, Physica A {\bf{357}}, 189 (2005).

\bibitem{barth00}
J. Barth, Surf. Sci. Rep. {\bf{40}}, 75 (2000).

\bibitem{sengupta05}
A. Sengupta, S. Sengupta, and G.I. Menon, Europhys. Lett. {\bf{70}}, 635 (2005).

\bibitem{wei00}
Q.-H. Wei, C. Bechinger, and P. Leiderer, Science {\bf 287}, 625 (2000).

\bibitem{hoefling06}
F. H\"ofling, T. Franosch, E. Frey, Phys. Rev. Lett. {\bf 96}, 165901 (2006).

\bibitem{dickson96}
R.M. Dickson, D.J. Norris, Y.-L. Tzeng, and W.E. Moerner, Science {\bf{274}}, 966 (1996).

\bibitem{weiss04}
M. Weiss, M. Elsner, F. Kartberg, and T. Nilsson, Biophys. J. {\bf{87}}, 3518 (2004).

\bibitem{hoefling13}
F. H{\"o}fling and T. Franosch, Rep. Prog. Phys. {\bf{76}}, 046602 (2013).

\bibitem{tolic04}
I.M. Toli{\'c}-N{\o}rrelykke, E.-L. Munteanu, G. Thon, L. Oddershede, and K. Berg-S{\o}rensen, Phys. Rev. Lett. {\bf{93}}, 078102 (2004).

\bibitem{chen00}
L. Chen, M. Falcioni, and M.W. Deem, J. Phys. Chem. B {\bf{104}}, 6033 (2000).

\bibitem{bystroem50}
A. Bytr{\"o}m and A.M. Bystr{\"o}m, Acta Cryst. {\bf{3}}, 146 (1950).

\bibitem{heuer05}
A. Heuer, S. Murugavel, and B. Roling, Phys. Rev. B {\bf{72}}, 174304 (2005).

\bibitem{indrani94}
A.V. Indrani and S. Ramaswamy, Phys. Rev. Lett. {\bf 73}, 360 (1994).

\bibitem{goetze99}
W. G\"otze, J. Phys.: Condens. Matter {\bf{11}}, A1 (1999).

\bibitem{debenedetti01}
P.G. Debenedetti and F.H. Stillinger, Nature (London) {\bf{410}}, 259 (2001).

\bibitem{angell95}
C.A. Angell, Science {\bf{267}}, 1924 (1995).

\bibitem{sciortino05}
F. Sciortino, J. Stat. Mech. P05015 (2005).

\bibitem{heuer08}
A. Heuer, J. Phys.: Condens. Matter {\bf{20}}, 373101 (2008).

\bibitem{durbin96}
S.D. Durbin and G. Feher, Annu. Rev. Phys. Chem. {\bf{47}}, 171 (1996).

\bibitem{dill97}
K.A. Dill and H.S. Chan, Nat. Struct. Mol. Biol. {\bf{4}}, 10 (1997).

\bibitem{hyeon03}
C. Hyeon and D. Thirumalai, Proc. Nat. Acad. Sci. {\bf{100}}, 10249 (2003).

\bibitem{janovjak07}
H. Janovjak, H, Knaus, and D.J. Muller, J. Am. Chem. Soc. {\bf{129}}, 246 (2007).

\bibitem{wolynes92}
P.G. Wolynes, Acc. Chem. Res. {\bf{25}}, 513 (1992).

\bibitem{jenkins08}
M.C. Jenkins and S.U. Egelhaaf, Adv. Colloid Interface Sci. {\bf 136}, 65 (2008).

\bibitem{ashkin97}
A. Ashkin, Proc. Natl. Acad. Sci. {\bf{94}}, 4853 (1997).

\bibitem{molloy02}
J.E. Molloy and M.J. Padgett, Contemp. Phys. {\bf{43}}, 241 (2002).

\bibitem{hanes09}
R.D.L. Hanes, M.C. Jenkins and S.U. Egelhaaf, Rev. Sci. Instrum. {\bf 80}, 083703 (2009).

\bibitem{crocker96}
J.C. Crocker and D.G. Grier, J. Colloid Interface Sci. {\bf 179}, 298 (1996).

\bibitem{haenggi90}
P. H\"anggi, P. Talkner, and M. Borkovec, Rev. Mod. Phys. {\bf{62}}, 251 (1990).


\end{thebibliography}
\end{document}